\documentclass[twocolumn]{emulateapj}

\usepackage{xcolor}

\shorttitle{The Braking Index of Millisecond Magnetars} 
\shortauthors{Lasky et al.}

\begin{document}

\title{The braking index of millisecond magnetars}

\author{Paul D. Lasky}
\affiliation{Monash Centre for Astrophysics, School of Physics and Astronomy, Monash University, Victoria 3800, Australia}
\affiliation{OzGrav: The ARC Centre of Excellence for Gravitational-wave Discovery, Hawthorn, Victoria 3122, Australia }
\email{paul.lasky@monash.edu}

\author{Cristiano Leris}
\affiliation{Monash Centre for Astrophysics, School of Physics and Astronomy, Monash University, Victoria 3800, Australia}

\author{Antonia Rowlinson}
\affiliation{Anton Pannekoek Institute, University of Amsterdam, Postbus 94249, NL- 1090 GE, Amsterdam, the Netherlands}
\affiliation{Netherlands Institute for Radio Astronomy (ASTRON), PO Box 2, NL-7990 AA Dwingeloo, the Netherlands}

\author{Kostas Glampedakis}
\affiliation{Departamento de F\'isica, Universidad de Murcia, Murcia, E-30100, Spain}
\affiliation{Theoretical Astrophysics, University of T\"ubingen, Auf der Morgenstelle 10, T\"ubingen, D-72076, Germany}

\begin{abstract}
We make the first measurement of the braking index $n$ of two putative millisecond magnetars born in short gamma-ray bursts.  We measure $n=2.9\pm0.1$ and $n=2.6\pm0.1$ for millisecond magnetars born in GRB 130603B and GRB 140903A respectively.  The neutron star born in GRB 130603B has the only known braking index consistent with the fiducial $n=3$ value.  This value is ruled out with 99.95\% confidence for GRB 140903A.  We discuss possible causes of $n<3$ braking indices in millisecond magnetars, showing that several models can account for the measurement of the braking index in GRB 140903A, while it is more difficult to account for a braking index consistent with $n=3$. 
\end{abstract}


\section{Introduction}

Observations of gamma-ray bursts and superluminous supernovae show evidence for ongoing energy injection following the prompt emission \citep{nousek06,obrien06,zhang06}, which is commonly attributed to the birth of rapidly rotating, highly magnetised neutron stars, known as a millisecond magnetars \citep[e.g.,][]{rowlinson13,lu14,inserra16}.  The spindown of the nascent neutron star drives high-energy emissions that are observed as long-lasting ($\gtrsim10^3\,\mbox{s}$) X-ray plateaus.  

The fiducial millisecond magnetar model relates the evolution of the star's spin frequency $\Omega(t)$ to the X-ray light curve \citep{zhang01,metzger11}. The original model assumes that the rapidly rotating star loses angular momentum through a combination of gravitational waves and electromagnetic radiation, although the amount of energy lost to gravitational-wave emission is small compared to electromagnetic losses \citep{lasky16,ho16,moriya16}. In general, the spindown of a neutron star can be described by the torque equation
\begin{equation}
	\dot{\Omega}=-k\Omega^n,\label{eq:spindown}
\end{equation}
where $k$ is a constant of proportionality and $n$ is the braking index.

An unchanging, dipolar magnetic field in vacuo implies a theoretical braking index of $n=3$ \citep{ostriker69}.  This fiducial assumption is built into the millisecond magnetar model, and leads to a prediction that the light curve luminosity decays as $L\propto t^{-2}$ at late times \citep{zhang01}.  Gamma-ray burst and superlumnious supernova light curves are usually fit assuming a braking index of $n=3$ \citep[e.g.,][]{troja07,rowlinson13,chatzopoulos13}.

In principle, Eq. (\ref{eq:spindown}) should equally apply to the spindown of rotation-powered pulsars.  Empirically though, not a single pulsar with a measured braking index is consistent with $n=3$, with all but one falling below $n\lesssim3$ \citep[see][and references therein]{archibald16,clark16,marshall16}.  More realistic calculations of pulsars and their magnetospheres ubiquitously predict $n\lesssim3$ \citep[e.g.,][]{melatos97}.

In this Letter, we make the first measurement of the braking index of two millisecond magnetars.  In particular, short gamma-ray bursts GRB 130603B and GRB 140903A, which were both observed with the {\it Swift} telescope and subsequently with XMM and Chandra respectively.  These late-time observations ($\gtrsim10^5\,\mbox{s}$ after the initial burst) allow us to make accurate measurements of the power-law decay of the light curve, and hence get tight constraints on the braking indices for the millisecond magnetars \footnote{The only other short GRB with such late time observations, albeit with only \textit{Swift}, is GRB 051221A, however \citet{lu15} claim the temporal and spectral properties pre and post break are consistent with an external forward shock, with only the plateau phase being due to continuous energy injection.}

We find that the braking index for the millisecond magnetar born in GRB 130603B is $n=2.9\pm0.1$ (1$\sigma$ confidence level), and hence consistent with $n=3$.  On the other hand, the millisecond magnetar born in GRB 140903A has $n=2.6\pm0.1$, ruling out $n=3$ with 99.95\% confidence.  We discuss physical mechanisms that can cause sub-three braking indices, finding that these naturally arise from physically-realistic models of post-merger remnants.

\section{Observations and Model}
\subsection{Generalised millisecond magnetar model}\label{sub:theory}
As a neutron star spins down, rotational kinetic energy is lost from the system, $E=\frac{1}{2}I\Omega^2$, where $I$ is the star's moment of inertia.  The time derivative of this equation gives the rate of change of energy loss; a certain fraction of which is converted into X rays.  The X-ray luminosity is therefore $L=-\eta\dot{E}=-\eta\Omega\dot{\Omega}$, where $\eta$ is the efficiency in converting spin-down energy into X rays.  We assume throughout that $\eta$ is not a function of time; a point we discuss further below.  Integrating Eq.~(\ref{eq:spindown}) gives the evolution of $\Omega(t)$, implying the luminosity is
\begin{equation}
	L(t)=L_0\left(1+\frac{t}{\tau}\right)^{\frac{1+n}{1-n}}.\label{eq:luminosity}
\end{equation}
Here, $L_0\equiv\eta Ik\Omega_0^{1+n}$ is the initial luminosity, $\Omega_0\equiv\Omega(t=0)$ and $\tau\equiv\Omega_0^{1-n}/[(n-1)k]$ is the spindown timescale of the system.

Equation (\ref{eq:luminosity}) shows the characteristic plateau $L=L_0$ behaviour for early times $t\ll\tau$, and a power-law decay $L\propto t^{(1+n)/(1-n)}$ for $t\gg\tau$.  When $n=3$, Eq. (\ref{eq:luminosity}) recovers the familiar late-time $L\propto t^{-2}$ behaviour where $\tau$ is the electromagnetic spindown timescale \citep{zhang01}.  In this limit, the spindown timescale becomes the familiar electromagnetic spindown timescale $\tau=\tau_{\rm em}\equiv3c^3I/(B_p^2R^6\Omega_0^2)$, where $B_p$ is the dipole, poloidal component of the star's magnetic field, and $R$ is the stellar radius. Normalising to typical millisecond magnetar parameters \citep{lasky14},
\begin{equation}
\tau_{\rm em} \approx 5 \times 10^{3}  \left ( \frac{B_p}{10^{15}\,\mbox{G}} \right )^{-2} \left ( \frac{P}{1\,\mbox{ms}} \right )^2\, \mbox{s},\label{eq:tauem}
\end{equation}
where $P$ is the spin period. 

It is worth noting that Eq.~(\ref{eq:luminosity}) is different to the `standard' derivation in the literature for when the spindown is dominated by gravitational-wave emission.  In that case, the braking index is $n=5$, implying from Eq.~(\ref{eq:luminosity}) that the luminosity decays as $t^{-3/2}$, instead of the oft-quoted $t^{-1}$ \cite[e.g.,][]{zhang01,lasky16}.  The derivation of $t^{-1}$ assumes that only the electromagnetic dipolar component of the spindown energy contributes to the X-ray light curve, whereas here the only assumption that has been made is that some fixed fraction $\eta$ of the spindown energy is converted into X rays.

In this Letter, we fit Eq.~(\ref{eq:luminosity}), combined with an initial power-law decay $L=At^{-\alpha}$ describing the transition between the prompt emission and the plateau phase \citep{rowlinson13}, to the data (described below).  We use Bayesian nested sampling, which provides us with joint posterior probability densities for $\{L_0,\,\tau,\,n,\,A,\,\alpha\}$.

\subsection{GRB 130603B}\label{sec:GRBthirteen}
The short-duration GRB 130603B  generated much interest as it was the first credible detection of a kilonova associated with a short GRB \citep{berger13,tanvir13}.  The initial burst \citep{melandri13} was picked up by the Burst Alert Telescope \citep[BAT;][]{barthelmy05a} onboard \textit{Swift} with a duration of $T_{90}=0.18\pm0.02\,{\rm s}$ in the 15--350 keV band \citep{barthelmy13}.  The X-ray Telescope \citep[XRT;][]{burrows05} onboard \textit{Swift} detected a corresponding fading X-ray source $59\,{\rm s}$ after the initial burst \citep{kennea13}.  A late-time excess was also observed with XMM-Newton $\approx2.7$ and $\approx6.5$ days after the initial burst \citep{fong14}.

The millisecond magnetar model has been invoked to explain both XRT and XMM X-ray excesses \citep{fan13a,fong14,deugartepostigo14}.  These papers all used the fiducial magnetar model with a braking index of $n=3$, allowing the magnetic field and initial spin period to be measured.  

Here we fit the more general magnetar model to the same data as that of \citet{fong14,deugartepostigo14}, allowing for a variable breaking index.  The top panel of Fig.~\ref{fig:lightcurve} shows the XRT and XMM data, together with our fit using Eq.~(\ref{eq:luminosity}) and an initial power law that fits the prompt emission.  The solid blue curve shows the maximum-likelihood model, while the dark red band is the superposition of many light curve models, where each model is drawn from a single posterior sample.  

In Fig.~\ref{fig:corner} we show a corner plot of the posterior probability distributions for the parameters in the magnetar model; the red contours show the posterior distributions for GRB 130603B.  In Fig.~\ref{fig:brakingindex} we plot the one-dimensional marginalized posterior distribution for the braking index, $n$.  The red curve representing the braking index for GRB 130603B shows consistency with the fiducial $n=3$ braking index, with $n=2.9\pm0.1$, where the uncertainties correspond to one-sigma confidence intervals.

\subsection{GRB 140903A}\label{sec:GRBfourteen}
GRB 140903A triggered BAT on Sept. 3, 2014 \citep{cummings14,palmer14}, with XRT observations of the GRB field 74 s after the BAT trigger \citep{dapasquale14}.  Two observations with the \textit{Chandra X-ray Observatory} were taken $\approx3$ and  $\approx15$ days following the initial BAT trigger, respectively \citep{troja16}.    The X-ray and other multi-wavelength observations of the GRB afterglow have been used to determine an achromatic jet-break, and hence infer the existence of a jet with a narrow opening angle of $\theta\approx5^\circ$ \citep{troja16,zhang17}.  Moreover, the \textit{Swift} and \textit{Chandra} X-ray plateau and power-law decay have been well-modelled within the fiducial $n=3$ magnetar model \citep{zhang17}.

\begin{figure}
\plotone{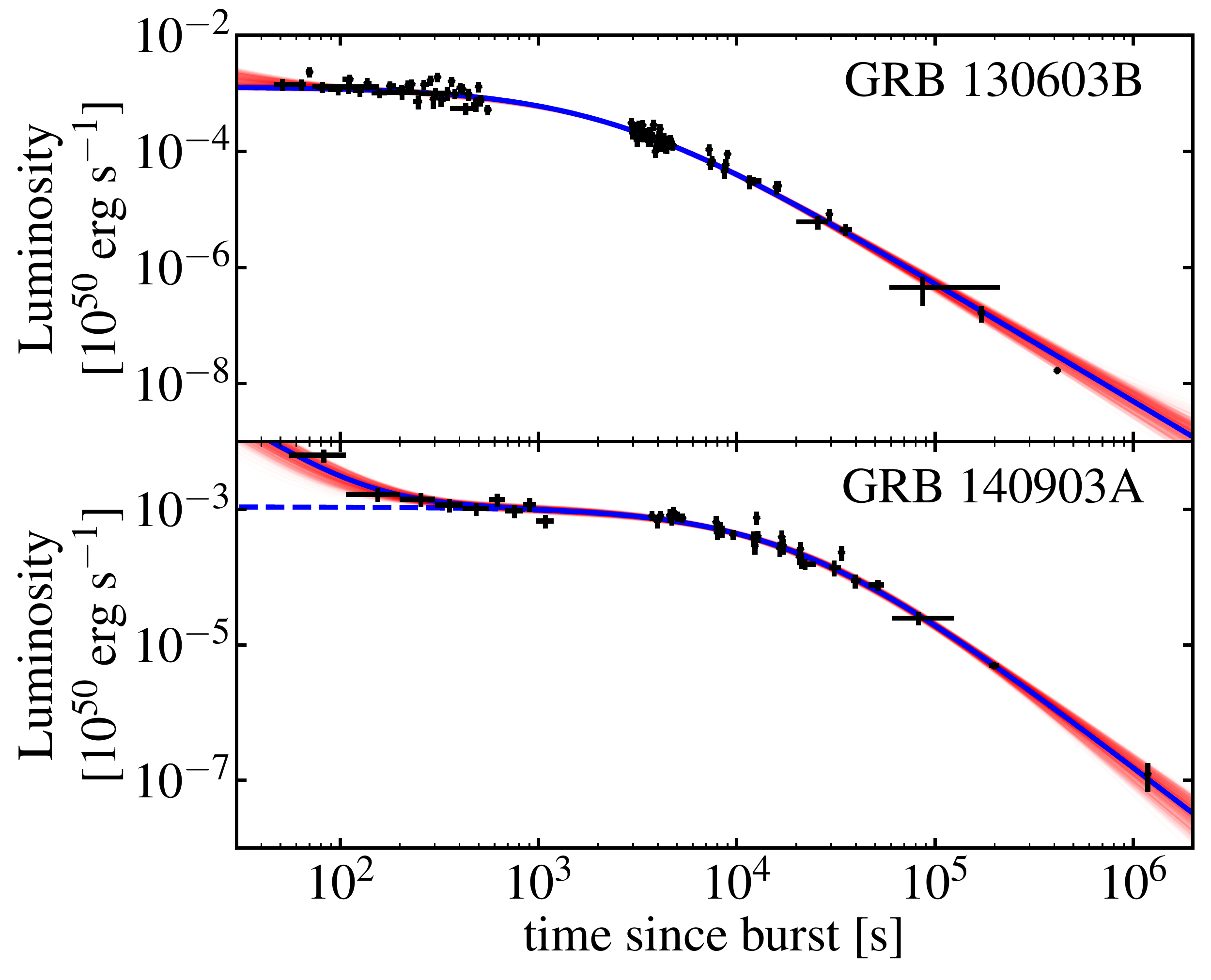}
\caption{\label{fig:lightcurve}X-ray lightcurves for GRB 130603B (top panel) and GRB 140903A (bottom panel).  In each panel, the black points represent the data (see sections~\ref{sec:GRBthirteen} and~\ref{sec:GRBfourteen} for details). The solid blue curve is the best-fit millisecond magnetar model, where the braking index ($n$; see Eq.~\ref{eq:luminosity}) is included in the fit.  The dashed blue curve is the best-fit millisecond magnetar model not including the contribution from the initial power-law decay.   The dark red band is the superposition of many light curve models, where each curve is drawn from a single posterior sample.
}
\end{figure}

We again fit the more general magnetar model to the same X-ray data used in \citet{troja16,zhang17}.  The bottom panel of Fig.~\ref{fig:lightcurve} shows the XRT and late-time \textit{Chandra} observations, together with our fit.  The solid blue curve again shows the maximum-likelihood model, and the dark red band shows the superposition of many light curves, each drawn from single posterior samples.    

\begin{figure}
\plotone{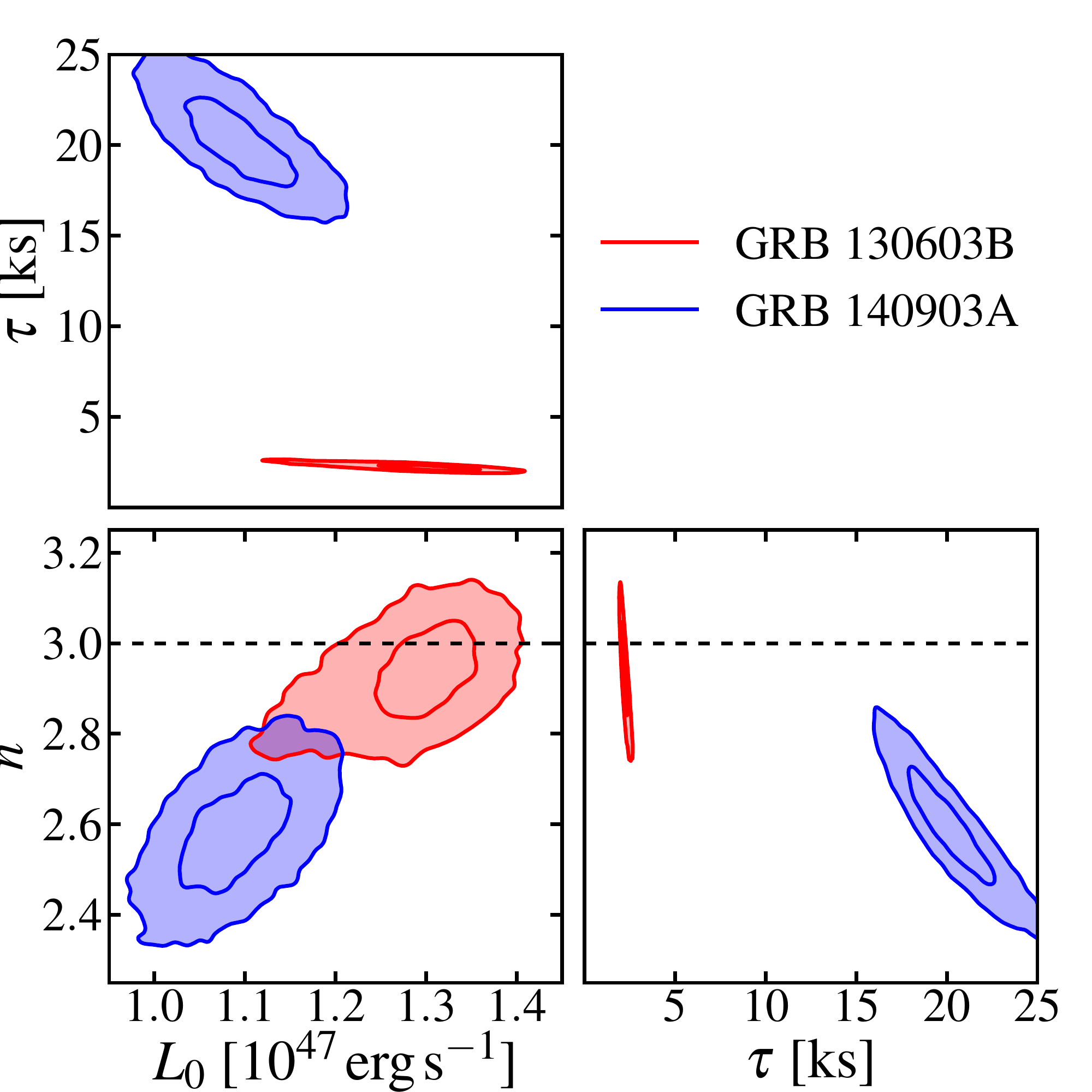}
\caption{\label{fig:corner}Posterior probability distributions for the parameters in Eq.~(\ref{eq:luminosity}) for GRB 130603B (red) and GRB 140903A (blue). The contours show the one- and two-sigma confidence intervals, and the dashed line indicates the fiducial value of $n=3$.}
\end{figure}

\begin{figure}
\plotone{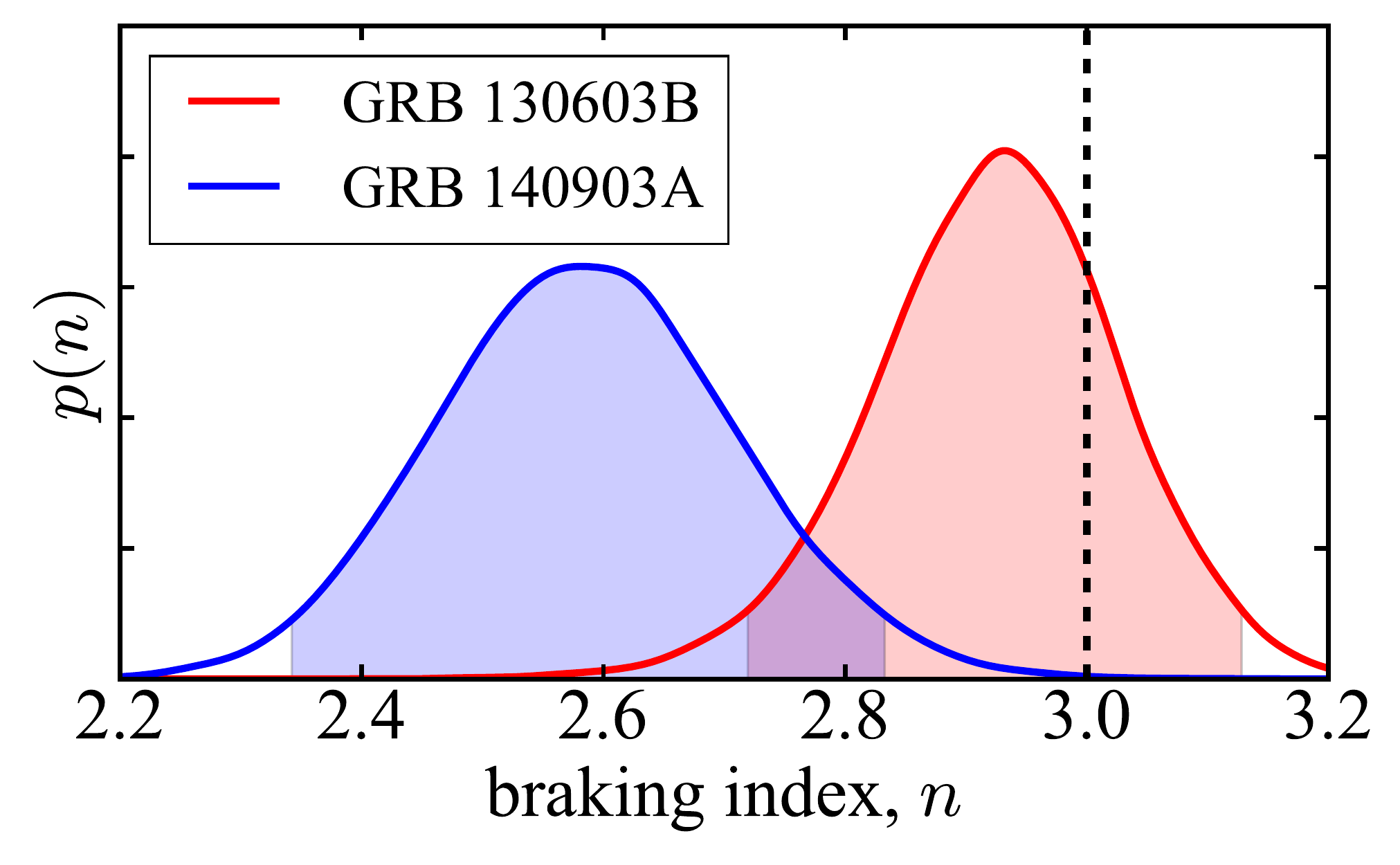}
\caption{\label{fig:brakingindex} One dimensional marginalized posterior distributions for the braking index $n$ for GRB 130603B (red) and GRB 140903A (blue).  The shaded regions show the two-sigma confidence intervals, and the dashed black line indicates the fiducial value of $n=3$.}
\end{figure}

The posterior probability distributions for the parameters of the millisecond magnetar model are shown as the blue contours in Fig.~\ref{fig:corner}.  The one-dimensional marginalized posterior distribution for the braking index is shown in blue in Fig.~\ref{fig:brakingindex}, which gives $n=2.6\pm0.1$.  The fiducial value of $n=3$ is ruled out with 99.95\% confidence.

\subsection{Comparison with pulsars}
\begin{figure}
\includegraphics[width=1.0\columnwidth]{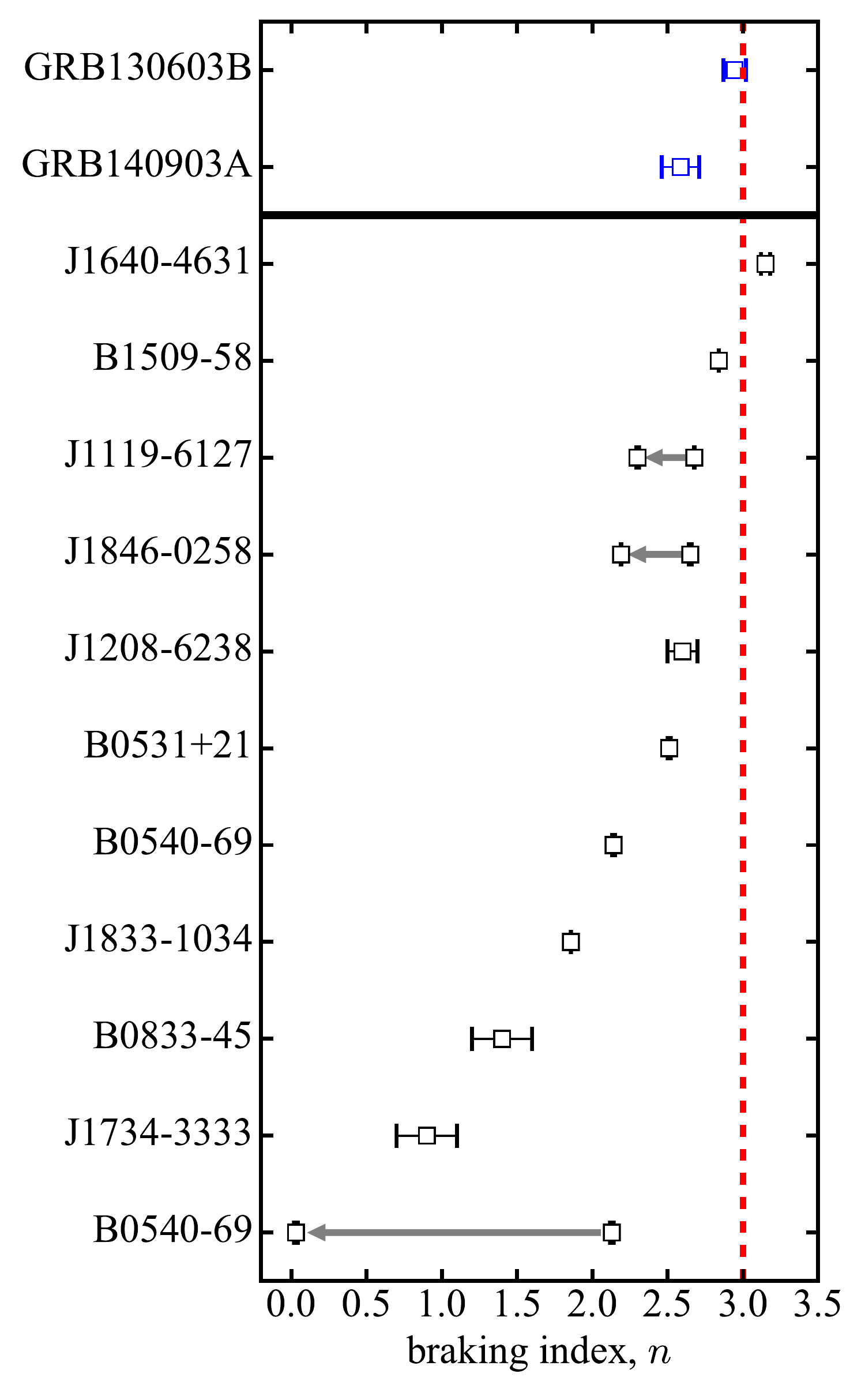}
	\caption{\label{fig:all_braking} Measured millisecond magnetar and pulsar braking indices.  The pulsar braking indices data are taken from \citet{archibald16,clark16,marshall16}, and references therein. Grey arrows represent changing braking indices seen in single pulsars between different spin-down states.}
\end{figure}

In Fig.~\ref{fig:all_braking} we plot the braking indices for all known pulsars where the long-term spindown is believed to be electromagnetically dominated \citep[see][and references therein]{archibald16,clark16}.  For comparison, we also plot the braking indices of the two neutron stars purportedly born in GRB 130603B and GRB 140903A.  The range of braking indices for pulsars spreads between $1\lesssim n\lesssim3.15$.  Clearly, there are not enough data to determine whether the sample of GRB-braking indices are statistically consistent with the distribution of pulsar braking indices with high confidence; we leave this as a topic for future work.

Intriguingly, Fig.~\ref{fig:all_braking} shows that the neutron star born in GRB 130603B has the only known braking index consistent with the fiducial $n=3$ value.  As we discuss below, it is relatively simple to devise models that explain $n\lesssim3$; the question therefore becomes: what is unique about the neutron star in GRB 130603B that makes it consistent with $n=3$?

%%%%%%%%%%%%%%%%%%%%%%%%%%%%%%%

\section{Theoretical explanation}\label{sec:theory}

The fiducial $n=3$ braking index is the one associated with the classic oblique rotator model in vacuo \citep{ostriker69}. 
The markedly more realistic model of \citet{goldreich69}, based on a charged-filled and force-free magnetosphere, also leads to the 
same prediction \citep[see][for a review]{spitkovsky08}. The departure of observed pulsar braking indices from the fiducial value 
can therefore be taken as evidence of additional physical processes affecting the spin-down of these objects. 

The literature posits an assortment of models that could explain the observed anomalous $n\lesssim3$ braking indices, 
but it is an open question as to which of those are relevant for the timescales, magnetic field strengths, and 
environmental conditions that are being discussed here. Here we focus on the various possibilities that could be 
of relevance for a millisecond magnetar system such as GRB 140903A.  And as we discuss, these may 
\emph{not} be the same mechanisms usually invoked for explaining anomalous braking indices of known pulsars.

%%%%%%%%%%%%%%%%%%%%%%%

\subsection{Modified magnetosphere spin-down}\label{sec:floatingfield}

A millisecond magnetar formed in the aftermath of a binary neutron star merger is likely endowed with a very strong toroidal magnetic field  
$\gtrsim10^{15}\,\mbox{G}$. This arises as both differential rotation of the body and the magneto-rotational instability work in concert to
amplify the field and wind up its lines;  these processes are expected to be present in the first $\sim 10-100\,\mbox{ms}$ following 
the merger \citep[e.g.,][]{rezzolla11,kiuchi14}. The induced magnetic field eventually quenches differential rotation on an Alfv\'en timescale $\ll1\,{\rm s}$ \citep{baumgarte00,shapiro00}.  
Therefore, after the initially chaotic period, the star settles to a rapidly, but rigidly rotating fluid ball with a strong, internal magnetic field. 
  
The generated field is dynamically unstable and rapidly rearranges itself to a state of hydromagnetic 
equilibrium where both poloidal and toroidal components are of comparable strength \citep[e.g.,][]{braithwaite09}. 
The global magnetic field rearrangement is likely to involve the bubbling up of toroidal magnetic flux to the stellar surface 
and into the magnetosphere \citep[e.g.,][]{kiuchi11}. The system therefore acquires a \emph{twisted} magnetosphere consisting of 
a strong mixed poloidal-toroidal field. This type of magnetosphere, originally modelled by \citet{thompson02}, is also believed to form in 
garden-variety magnetars \citep[see][and references therein]{turolla15}.  
Such twisted magnetospheres increase the spin-down torque in comparison to orthogonal vacuum dipoles, implying 
reduced values of the braking index; i.e., $n\lesssim3$ \citep{thompson02}. The amount of reduction is not unique, 
but largely depends on the field's radial profile and the degree of twist, with higher twist leading to a smaller braking index. 

An entirely different magnetospheric modification for producing an $n < 3$ spin-down has been proposed by \cite{contopoulos06}. 
In this model, which assumes a dipole force/twist-free magnetosphere, a distinction is made between the light-cylinder radius 
$R_{\rm lc} = c/\Omega$ (the cylindrical radius at which a magnetosphere rigidly corotating with the star would exceed the speed of light)  
and the separatrix radius  $R_{\rm c}$ between open and closed field lines. 

Although the standard assumption is that of $R_{\rm lc} =  R_{\rm c} $, this may not be a strictly-imposed physical necessity; 
$R_{\rm c}$ could lag behind $R_{\rm lc}$ if the spin-down is fast and the magnetic field line reconnection cannot keep up with 
the outward migrating light-cylinder  \citep[for details, see][]{contopoulos06}. Such a scenario could be strongly favoured in a 
millisecond magnetar, implying the spin-down torque is enhanced  as a result of the open field lines' larger aperture.  
A braking index of $ n<3$ therefore naturally emerges.

%%%%%%%%%%%%%%%%%%%%%%%%

\subsection{Magnetic axis evolution and other mechanisms}\label{sec:Magaxis}
Another possible explanation for $n\lesssim3$ braking indices relates to the evolution of the angle $\alpha$ between the star's 
rotation axis and its surface dipole magnetic field axis.  

This mechanism, for example, has been invoked to explain the braking index of the Crab pulsar \citep[][]{lyne15}.  
Here, the $k$ term in Eq.~(\ref{eq:spindown}) is a function time, and the braking index can be approximated as 
$n=3+2\Omega\dot{\alpha}/(\dot{\Omega}\tan\alpha)$. 
Clearly, $\dot{\alpha}>0$ leads to a braking index $n<3$.  

The time evolution of $\alpha$ largely depends on whether the dipole field can be considered rigidly attached 
to the star's `body frame' -- i.e., the deformed shape induced by the strong toroidal component. 
According to standard oblique rotator theory \citep{goldreich70}, the electromagnetic torque due to the exterior field drives the 
symmetry axis of the deformation towards (away from) the spin axis on a spindown timescale 
$\tau_{\rm em}$ if its direction with respect to the dipole axis makes an angle smaller (larger) than $ \approx 55^{\rm o}$. 
Assuming a fixed relative orientation between the surface dipole and internal toroidal field symmetry axes, 
the desired $\dot{\alpha} > 0 $ situation arises provided the dipole axis is significantly misaligned ($\gtrsim 55^{\rm o}$)
with respect to the spin axis since the latter axis is expected to lie close to the toroidal field's symmetry axis. 

The evolution of the relative orientation between the spin and deformation axes also couples to the emitted gravitational
waves \citep{cutler01}. Unlike the previous case, however, gravitational radiation always drives the two 
axes towards alignment, on a timescale $\tau_{\rm gw} \approx 2\times 10^4 (\epsilon_B/10^{-3})^{-2} (P/1\,\mbox{ms})^4\, \mbox{s}$, 
where $\epsilon_{\rm B}$ is the magnetic field-induced stellar ellipticity.  This is long compared to the electromagnetic spindown timescale---cf., Eq.~(\ref{eq:tauem})---unless the ellipticity is $\epsilon\gtrsim10^{-2}$, which is all but ruled out for systems in which the ellipticity can be measured \citep{lasky16}.

A natural way to drive an $\dot{\alpha} >0$ evolution is through the so-called `spin-flip' instability \citep[e.g.,][]{cutler02}.
Here, the strong internal toroidal field causes the star to become a prolate spheroid in the first few seconds after birth. 
Such an arrangement is unstable and, under the action of internal dissipation (in the present case bulk viscosity), the system is driven towards 
a state where the spin and toroidal symmetry axes are mutually orthogonal. If the dipole field is assumed to be `locked'  to the toroidal
component then this orthogonalisation implies  $\dot{\alpha} >0$.

The spin-flip could be a viable mechanism for modifying the braking index provided its timescale is comparable to
the spin-down timescale, $\tau_{\rm sf} \sim \tau_{\rm em}$. For much of the relevant parameter space, however, 
 $\tau_{\rm sf}$ is likely to be much shorter than $\tau_{\rm em}$; see discussion around Fig.~1 in \citet{lasky16}. 
For the two timescales to become comparable the magnetic ellipticity must be substantial, $\epsilon_B \gtrsim 10^{-3}$.
For such a system the spin-flip timescale is minimised at a relatively high temperature but the cooling in that regime
is so rapid that the instability actually kicks in at a lower temperature where $\tau_{\rm sf}$ is significantly longer 
(and comparable to $\tau_{\rm em}$). In this scenario, millisecond magnetars harbouring magnetic fields that are wound up sufficiently large are expected to have a $n <3$ distribution. On the other hand, weaker fields in the core leading to smaller ellipticities may give rise to $n \approx 3$ magnetars.

Finally, it is worth pointing out that mechanisms that have been invoked to explain the anomalous braking indices of
radio pulsars are unlikely to be of relevance to the case at hand. The resurfacing of an initially ``buried" magnetic field 
due to fallback material is a relatively slow process \citep[dominated by the Ohmic-diffusion timescale $t\sim1-100\,{\rm kyr}$;][]{vigano12};
much longer than the timescales associated with short GRB remnants.  Moreover, those calculations were done in the context of core-collapse supernovae where there is more fallback material to bury the field in the first place.
Similarly, a gradual change in the stellar moment of inertia due to the onset of neutron superfluidity \citep{ho12} could only take place in systems significantly older and colder than the ones considered here.

%%%%%%%%%%%%%%%%%%%%%%%%

\section{Conclusion}
In this Letter, we make the first measurements of the braking index $n$ of putative millisecond magnetars born in short gamma-ray bursts.  Observations of X-ray plateaus following short-gamma ray bursts indicate the presence of ongoing energy injection, commonly attributed to the rotational evolution of a nascent neutron star.  We show that the power-law exponent of the late-time ($\gtrsim10^3\,{\rm s}$) decay of these curves can be directly related to the braking index; see Eq.~(\ref{eq:luminosity}).

We show that the braking index of the magnetar in GRB 140903A is inconsistent with the fiducial value of $n=3$ predicted for an unchanging, rotating dipolar magnetic field.  However, as we propose in Sec.~\ref{sec:theory}, there are a number of models that naturally explain this for millisecond magnetars.  These include the presence of twisted components of the magnetic field in the magnetosphere (Sec.~\ref{sec:floatingfield}) or evolution of the angle between the magnetic axis and the star's rotation axis (Sec.~\ref{sec:Magaxis}).  Another possibility is that the efficiency of converting spin-down energy into X rays, $\eta$, evolves as a function of time; assuming $d\eta/dt<0$, this would also lead to the inference of a sub-three braking index -- see section \ref{sub:theory}.

Perhaps what warrants more attention is the braking index for the millisecond magnetar born in GRB 130603B which, at $n=2.9\pm0.1$, is consistent with the fiducial $n=3$ value.  \textit{This is the only empirically-measured braking index consistent with $n=3$.}  All sophisticated models of neutron star magnetosphere's tend to predict sub-three values for the braking index, especially when one considers those models relevant for the evolution of nascent stars born from binary neutron star mergers (see Sec.~\ref{sec:theory} for a detailed discussion). It is worth mentioning that $\sim68\%$ of the marginalised posterior for the braking index of the neutron star in GRB 130603B predicts $n<3$.

It is tempting to read more into the X-ray light curve associated with GRB 130603B than we have done herein.  For example, the last XMM data point taken more than 6 days after the burst (see Fig.~\ref{fig:lightcurve}) lies below almost all of the light curves generated from the posterior samples.  It is therefore tempting to say that the braking index is actually evolving, and that one should include a $dn/dt$ term in the torque equation.  However, there is simply not enough data at late times to warrant such a hypothesis.  Clearly, to make such a claim one would want more data for times $t\gtrsim10^6\,{\rm s}$, which will not be forthcoming for this GRB.

In lieu of more data for this particular GRB, we are left with potential statistical analyses of GRB light curves to determine braking indices of the population of millisecond magnetars born in short GRBs.  We leave this to future work, although note that a majority of short GRBs detected with \textit{Swift} do not contain late-time observations from XMM or \textit{Chandra} as with the GRBs analysed herein.  Sensitivity limitations of \textit{Swift's} XRT limit the final data point to $t\lesssim10^5\,{\rm s}$ after the prompt emission, implying constraints on the braking index of magnetars born in such GRBs will come with commensurately larger uncertainties.
 
The analysis herein can also be extended to light curve analyses of millisecond magnetars born in long GRBs and superluminous supernovae.  Such analyses, however, are fraught with more difficulties than presented herein.  For example, both long GRBs and superluminous supernovae likely have denser and messier environments surrounding the initial explosion, implying extra complications in the spin-down torque from, for example, fallback accretion onto the newborn neutron star.

\acknowledgements
PDL is grateful to Bing Zhang for useful conversations.  We thank the referee for providing thorough feedback.  PDL is supported by ARC Future Fellowship FT160100112 and CoE CE170100004.  PDL and KG are supported by NewCompstar (a COST-funded Research Networking Programme).

\bibliographystyle{aasjournal}
%\bibliography{PlateauBib}

\label{lastpage}

\end{document}